\documentclass[aps,prl,twocolumn,showpacs,superscriptaddress,groupedaddress]{revtex4}  
\usepackage{graphicx}  
\usepackage{dcolumn}   
\usepackage{bm}        
\usepackage{amssymb}   
\usepackage{slashed}
\usepackage{color}
\usepackage{lipsum}

\begin{document}

\widetext

\title{Inclusive approach to hunt for the beauty-charmed baryons $\Xi_{bc}$}

\author{Qin~Qin}\email{qqin@hust.edu.cn}
\affiliation{School of Physics, Huazhong University of Science and Technology, Wuhan 430074, China}
\author{Yu-Ji Shi}\email{shiyuji92@126.com}
\affiliation{Helmholtz-Institut f\"ur Strahlen- und Kernphysik and Bethe Center for Theoretical Physics,
  Universit\"at Bonn, 53115 Bonn, Germany}
\author{Wei Wang}
\affiliation{MOE Key Laboratory for Particle Physics, Astrophysics and Cosmology,  Shanghai Key Laboratory for Particle Physics and Cosmology, School of Physics and Astronomy, Shanghai Jiao Tong University, Shanghai, 200240, China}
\author{Guo-He Yang}
\affiliation{School of Physics, Huazhong University of Science and Technology, Wuhan 430074, China}
\author{Fu-Sheng Yu}
\affiliation{School of Nuclear Science and Technology, 
Lanzhou University, Lanzhou 730000, China}
\affiliation{Center for High Energy Physics, Peking University, Beijing 100871, China}
\author{Ruilin Zhu\footnote{corresponding authors: Qin Qin, Yu-Ji Shi and Ruilin Zhu}}\email{rlzhu@njnu.edu.cn}
\affiliation{Department of Physics and Institute of Theoretical Physics,
Nanjing Normal University, Nanjing, Jiangsu 210023, China}

\begin{abstract}
With a distinctive internal structure from all established hadrons, the beauty-charmed baryons $\Xi_{bc}$ 
can provide us with new points of view to decipher the strong interaction.
In this work, we point out that the inclusive $\Xi_{bc} \to \Xi_{cc}^{++}+X$ decay is a golden channel for the experimental discovery 
of $\Xi_{bc}$ at the LHC. A unique feature of this process is that the $\Xi_{cc}^{++}$ is displaced, which greatly reduces the combinatorial background. 
A feasibility analysis is performed on the $\Xi_{bc}^+$ search, which is expected to have a longer lifetime than $\Xi_{bc}^0$ and thus a better displacement resolution. 
The $\Xi_{bc}^+ \to \Xi_{cc}^{++}+X$ branching ratio is calculated within the heavy diquark effective theory. 
Combining the $\Xi_{bc}$ production rate 
and the $\Xi_{cc}^{++}$ detection efficiency, we anticipate that hundreds of signal events will be collected during LHCb Run 3.
\end{abstract}


\maketitle


\textit{Introduction.} ---
Doubly heavy baryons, especially the  $\Xi_{cc}^{++}$,  have recently  received much attention~\cite{LHCb:2017iph}. 
Different from other baryons with one or zero heavy quarks, doubly heavy baryons resemble a `double-star' core surrounded by a light `planet'.
The $\Xi_{cc}^{++}$ discovery also motivated  studies to  probe the nature of exotic four quark states or structures, {\it e.g.} cusps or 
true resonances (see 
{\it e.g.}~\cite{Karliner:2017qjm,Eichten:2017ffp,Liu:2019zoy}). Very recently, a first doubly heavy tetraquark candidate $T_{cc}^+$ was observed 
by the LHCb~\cite{LHCb:2021vvq,LHCb:2021auc}. However unlike in doubly charmed systems,  the `double-star' core in the beauty-charm baryons $\Xi_{bc}$ is imbalanced,  resulting in diverse features. Compared to the charm-charm binary, the beauty-charm core is expected to have a smaller size, behaving more like a point particle. Moreover, the beauty-charmed baryons 
involve more energy scales, the beauty mass, the charm mass and the nonperturbative QCD scale $\Lambda_{\text{QCD}}$, so they implicate more affluent dynamics. Thereby, the beauty-charmed baryons would provide a unique new hadronic platform  to decode the strong interaction.

Experimentalists have made abundant efforts to search for the beauty-charmed baryons $\Xi_{bc}$. However, such searches 
are much more difficult than those for $\Xi_{cc}^{++}$.   For example, 
the exclusive channels $\Xi_{bc}^0\to D^0pK^-$~\cite{LHCb:2020iko} and $\Xi_{bc}^0\to\Xi_c^+\pi^-$~\cite{LHCb:2021xba} 
were used to search for $\Xi_{bc}$ at the LHCb, but no definitive evidence was established. 
With some theoretical and experimental inputs, the experimental upper limit on the 
$\Xi_{bc}^0\to\Xi_c^+\pi^-$ branching ratio can be extracted from Ref.~\cite{LHCb:2021xba} to be $\mathcal{O}(10^{-4})$. 
Comparing it to the theoretical prediction~\cite{Wang:2017mqp}, we find a big gap of about 3 orders of magnitude.
One difficulty  in such exclusive searches lies in the limited production rate for $\Xi_{bc}$ at the LHC, but 
a bigger challenge  is due to the very low reconstruction efficiency,
because a beauty typically decays with fractions of $\mathcal{O}(10^{-3})$ even to the most
abundant exclusive final states~\cite{Wang:2017mqp,Kiselev:2001fw,Han:2021gkl}. To overcome this difficulty,
we propose an approach to search for $\Xi_{bc}$ via an inclusive decay channel $\Xi_{bc}\to \Xi_{cc}^{++} +X$,
where $X$ stands for all possible particles.

\begin{figure}\begin{center}
\includegraphics[scale=0.1]{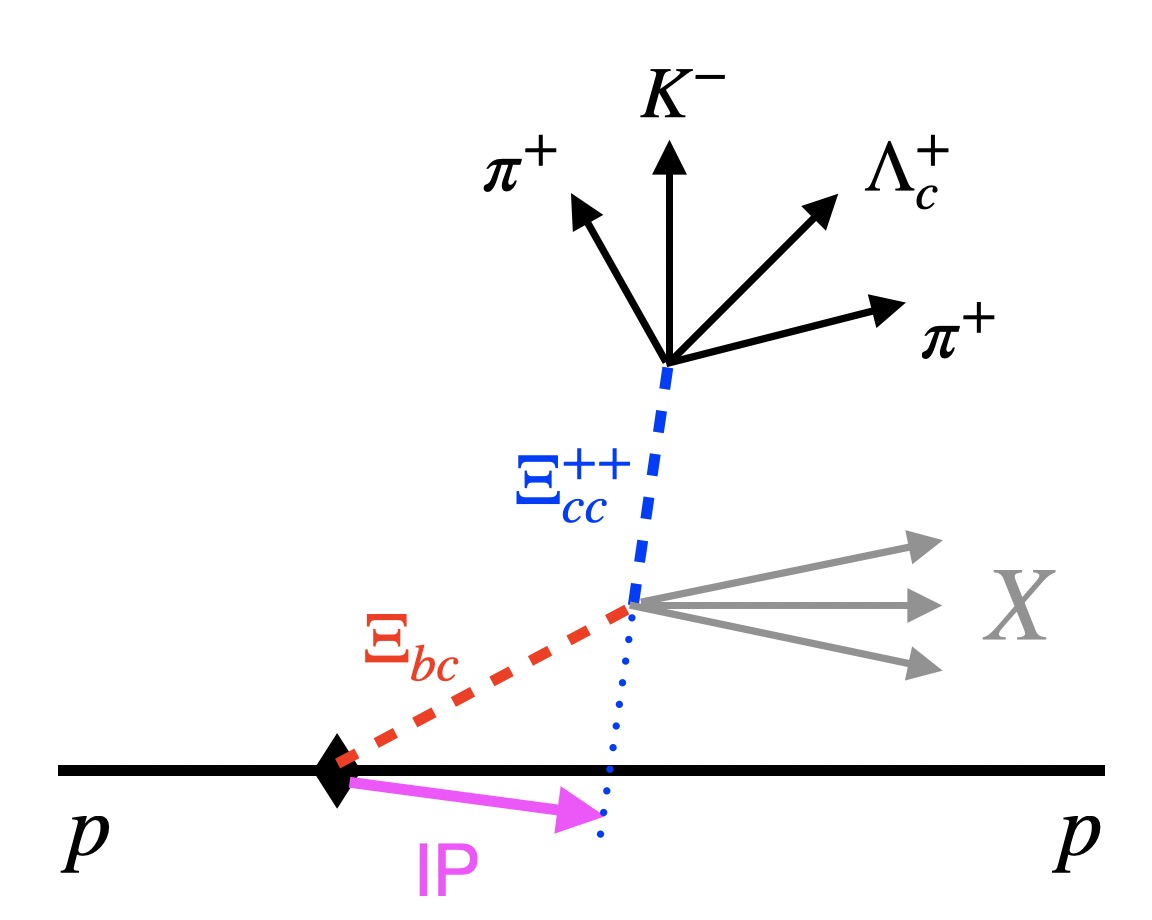}
\caption{Sketch of $\Xi_{bc}$ production and decay at the LHC. The secondary decay vertex of 
$\Xi_{bc}\to \Xi_{cc}^{++}+X$ is displaced from the proton-proton collision vertex, which produces a unique 
signal: a displaced $\Xi_{cc}^{++}$.} \label{fig:displace}
\end{center}
\end{figure}

This inclusive approach to search for $\Xi_{bc}$ has multiple  advantages.
Firstly, it has a much larger branching ratio than any exclusive decay channel. Secondly, the detection 
efficiency is greatly improved because only $\Xi_{cc}^{++}$ needs to be reconstructed. 
Lastly but very importantly, because the weakly decaying $\Xi_{bc}$ has a relatively long lifetime and can  typically form a sub-millimeter
displaced secondary decaying vertex, the $\Xi_{cc}^{++}$'s generated  from $\Xi_{bc}$ do not draw back
to the primary proton-proton collision vertices. This feature characterized by a nonzero impact parameter (IP)
can clearly distinguish the signal events from the main background, strongly produced $\Xi_{cc}^{++}$'s from the primary vertices. 
To clarify this point, a diagrammatic sketch is displayed in 
Fig.~\ref{fig:displace}. The use of the IP
has been applied at the LHCb for a long time (see {\it e.g.}~\cite{LHCb:2010wqx}), and it was proposed that 
displaced $B_c$ mesons can be used to search for $\Xi_{bb}$ in~\cite{Gershon:2018gda}, which greatly inspired the 
original idea of this work. 
According to~\cite{Gershon:2018gda}, the relatively long lifetime of $\Xi_{bc}^+$~\cite{Cheng:2019sxr} 
can ensure it a good IP resolution, while the situation will be worse for $\Xi_{bc}^0$.
Apart from the IP, the distance between the $\Xi_{cc}^{++}$ decay vertex and the 
primary collision vertex can also be used in the search to improve the sensitivity.

As the $\Xi_{bc}$ production rates and lifetimes have been evaluated in {\it e.g.}~\cite{Zhang:2011hi,Ali:2018xfq} 
and~\cite{Cheng:2019sxr} (see also references therein), respectively, the remaining key issue is the $\Xi_{bc}\to \Xi_{cc}^{++} +X$ branching ratio. 
Based on the heavy diquark effective theory \cite{Brambilla:2005yk,An:2018cln,Shi:2020qde}, we  
calculate the branching ratio and find $\mathcal{B}(\Xi_{bc}^+\to \Xi_{cc}^{++} +X)\approx$ 7\%. Moreover, because the signal mainly characterizes in a displaced 
$\Xi_{cc}^{++}$, the reconstruction efficiency of the signal events is close to the detection efficiency of $\Xi_{cc}^{++}$, 
which can be reliably extracted from previous experiments. Combining all this information, 
we find that hundreds of signal events are expected during LHCb Run 3, with an integrated 
luminosity of 23 fb$^{-1}$ by 2024. Consequently, the proposed 
inclusive approach for the $\Xi_{bc}$ search is most feasible and also timely for the LHCb study.

{\textit{Decay rate.}} ---
In the feasibility analysis of this approach, the inclusive $\Xi_{bc} \to \Xi_{cc}^{++}+X$
decay rate was calculated with the following several steps. Based on the heavy quark symmetry and the
heavy diquark symmetry, each step of the calculation is trustworthy. Firstly, under the heavy
(di)quark symmetry, it can be
demonstrated that the leading contribution to the inclusive $\Xi_{bc} \to \Xi_{cc}^{++}+X$ decay is from
$\mathcal{X}_{bc} \to \mathcal{X}_{cc} +\bar{f}f'$, where $\mathcal{X}_{QQ'}$ stands for a heavy diquark
constituted by the heavy $Q$ and $Q'$ quarks and $f^{(\prime)}$ can be any possible quarks or leptons.
Subsequently, the unknown $\mathcal{X}_{bc} \to \mathcal{X}_{cc}$ diquark transition current was evaluated
by matching from the $b\to c$ transition current. Afterwards, the decay rate of $\Xi_{bc} \to \Xi_{cc}^{++}+X$ was
numerically calculated, with possible theoretical uncertainties taken into account.

We first validate the treatment of  the two heavy quarks $QQ'$ as a point-like
object in a doubly heavy baryon.The $QQ'$ form a color anti-triplet
and have an attractive potential.  As illustrated by \cite{Brodsky:2011zs,Yan:2018zdt,Hu:2005gf},
the distance between the two heavy quarks is estimated as $r_{QQ}\sim 1/(m_Qv)$ with $v$ being the heavy quark velocity in the baryon rest frame,
while the spatial size of the light quark in the baryon is $r_{Qq}\sim 1/{\Lambda_{\text{QCD}}}$. 
Furthermore, it can be deduced that $v$ is small if $m_Q$ is heavy enough \cite{Bodwin:1994jh}.
Numerical calculations confirm the hierarchy by giving $v_c^2\sim 0.3$
and $v_b^2\sim 0.1$~\cite{Bodwin:1994jh,Quigg:1979vr}. It indicates that
$m_b v_b^2\sim m_c v_c^2\sim\Lambda_{\text{QCD}}$, so $r_{QQ}/r_{Qq}\sim \Lambda_{\text{QCD}}/(m_{Q}v) \ll 1$.
In conclusion, the two heavy quarks can be treated as a point-like diquark compared to the baryon size.
This greatly simplifies the structure of a three-quark system to a bound state of a heavy diquark and a light quark.

Benefitting from the quark-diquark picture, we can formulate the inclusive decay of a doubly heavy baryon 
within the heavy diquark effective theory~\cite{Brambilla:2005yk,An:2018cln,Shi:2020qde}. Performing the 
operator product expansion, the inclusive $\Xi_{bc} \to H_{cc}+X$
decay rate can be expanded in inverse powers of the diquark mass $M_\mathcal{X}$, with the leading-power contribution given
by the free diquark decay rate,
\begin{eqnarray}\label{eq:lp}
\Gamma(\Xi_{bc}\to H_{cc}+X)
= \sum_{f,f'}\Gamma(\mathcal{X}_{bc}\to \mathcal{X}_{cc}\bar{f}f') +\mathcal{O}\left({1\over M_{\mathcal{X}}} \right) .
\end{eqnarray}
The fermion pairs $\bar{f}f'$ include $\bar{v}_\ell\ell^-$ ($\ell = e,\mu,\tau$) and
$\bar{u}d,\bar{u}s,\bar{c}d,\bar{c}s$. 
The $H_{cc}$ represents all doubly charmed hadrons, including the ground-state baryons 
$\Xi_{cc}^{++(+)}$, $\Omega_{cc}^{+}$ and the tetraquarks $T_{cc}$, and their excited states as well. 
As the fragmentation rates to strange baryons and to tetraquarks are much smaller than those
to non-strange baryons (see, {e.g.},~\cite{LHCb:2014ofc,LHCb:2021qbv}), the fragmentations to 
$\Omega_{cc}$ and $T_{cc}$ and their excited states are neglected in the
following discussions. 
For the non-strange baryons, the excited states eventually decay strongly (or electromagnetically) 
into $\Xi_{cc}$. Therefore, all
the $\Xi_{bc} \to H_{cc}+X$ decay processes produce a displaced $\Xi_{cc}$, half
$\Xi_{cc}^{++}$ and half $\Xi_{cc}^+$ by the isospin symmetry, {\it i.e.},
$\Gamma(\Xi_{bc} \to H_{cc}+X) \approx\Gamma(\Xi_{bc} \to \Xi_{cc} + X) \approx 2 \Gamma(\Xi_{bc} \to \Xi_{cc}^{++} + X)$.

\begin{figure}\begin{center}
\includegraphics[scale=0.08]{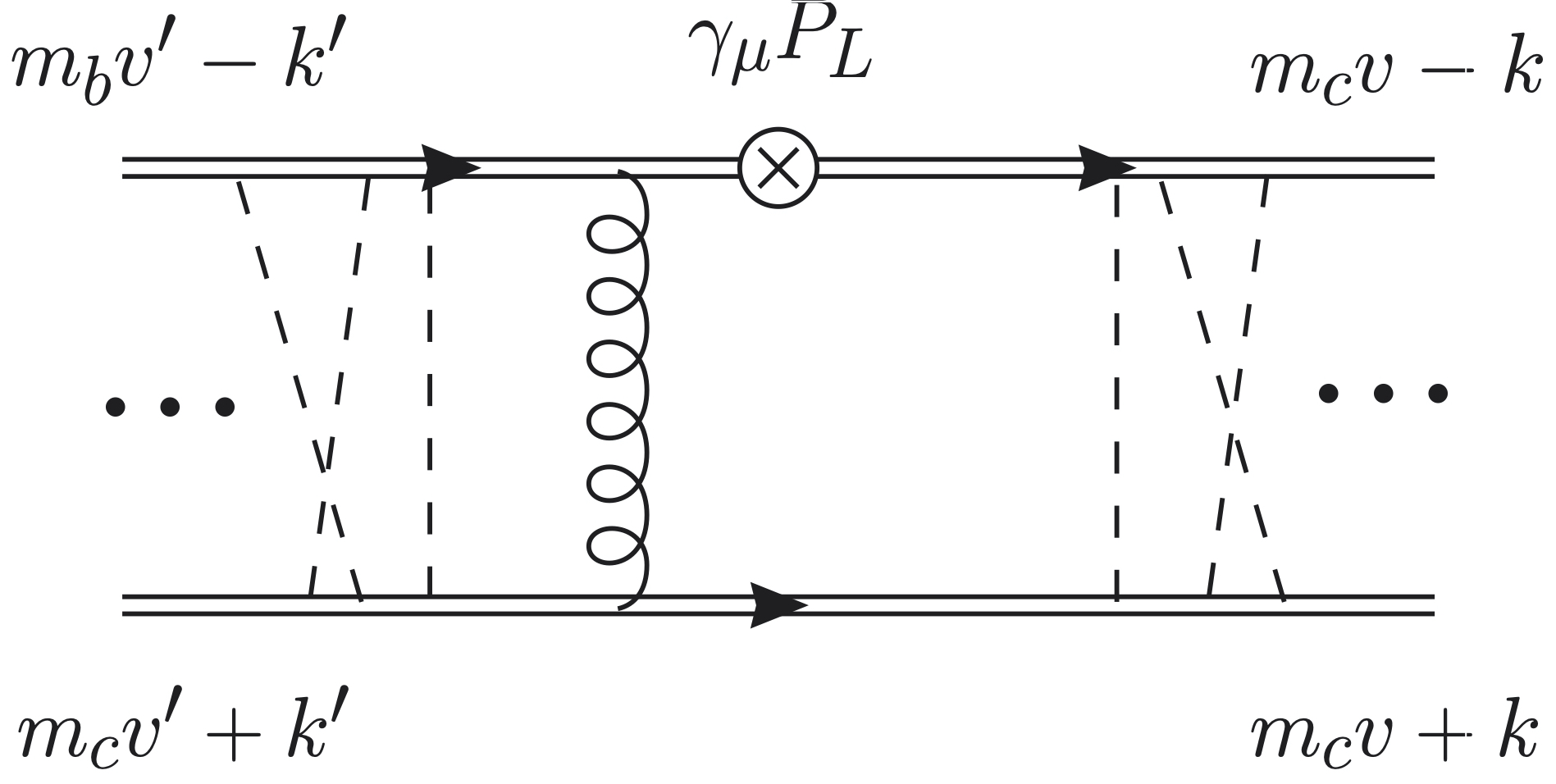}
\caption{A sample Feynman diagram of the $\mathcal{X}_{bc}\to \mathcal{X}_{cc}$ diquark transition induced by the $V-A$ current at large recoil. The double lines denote the heavy quarks. The gluon line close to the weak vertex denotes a hard gluon. The dashed lines denote any number of soft gluons which can be absorbed into the initial and final diquark wave functions. The velocity of $\mathcal{X}_{cc(bc)}$ is $v^{(\prime)}$. The relative momentum of the two heavy quarks in $\mathcal{X}_{cc(bc)}$ is $k^{(\prime)}$.} \label{fig:maxrecoil}
\end{center}
\end{figure}

In the evaluation of $\Gamma(\mathcal{X}_{bc}\to \mathcal{X}_{cc}\bar{f}f')$ induced by the weak interaction vertices
such as $\bar{c}\gamma^\mu P_L b \bar{f}\gamma_\mu P_L f'$ with the left-handed projector $P_L\equiv(1-\gamma_5)/2$, the $\bar{f}f'$ part can be
factorized out at the leading order of the strong coupling constant $\alpha_s$, and the key issue in the calculation is the remaining diquark current 
$\langle \mathcal{X}^i_{cc} |  \bar{c}\gamma^\mu P_Lb |\mathcal{X}^l_{bc} \rangle$,
where $i,l$ are color indices. The S-wave diquark $\mathcal{X}_{bc}$ is either a scalar or axial-vector,
but as implied by studies of the beauty-charmed baryon spectroscopy \cite{Ebert:1996ec,He:2004px,Yu:2018com,Weng:2018mmf,Li:2019ekr}, 
an axial-vector $\mathcal{X}_{bc}$ state is dominant in the $\Xi_{bc}$ baryons. On the other hand, $\mathcal{X}_{cc}$ can only be an axial-vector due to the flavor and spin symmetries.
The calculation of the diquark current is performed in two different kinematic
regions, the large recoil region and the small recoil region. In the former, perturbative calculation
is applicable because typically a hard gluon exchange between the spectator quark and the weak
interacting quarks is required, as displayed in FIG. \ref{fig:maxrecoil}. In practice, we adopt the non-relativistic
QCD (NRQCD) factorization for this calculation, which were applied to calculate the $B_c\to \eta_c, J/\psi$ form factors~\cite{Qiao:2012vt,Qiao:2012hp,Zhu:2017lqu}.
In the small recoil region, the so-called soft overlap contribution is dominant and the perturbative QCD expansion is less trustworthy. However, the heavy quark symmetry determines the form of the diquark current at the zero recoil point.
For the intermediate region, we use a simplified $z$-series expansion \cite{Bourrely:2008za} to perform the interpolation.

The vector and axial-vector diquark currents can be parametrized as 
\begin{eqnarray}\label{eq:amplitude}
&&\langle \mathcal{X}^i_{cc}(v,\epsilon) |  \bar{c}\gamma^\mu b |\mathcal{X}^l_{bc}(v',\epsilon') \rangle \nonumber \\
&= & \delta_{il} \sqrt{2M_{cc}M_{bc}} \Big[ - a_0\epsilon^*\cdot\epsilon' v^{\prime\mu}
- a_1 \epsilon^*\cdot\epsilon' v^\mu  \nonumber\\
&&+ a_2 \epsilon^*\cdot v' \epsilon^{\prime\mu}  + a_3 v\cdot\epsilon' \epsilon^{*\mu} \nonumber \Big] \; ,  \nonumber \\
&&\langle \mathcal{X}^i_{cc}(v,\epsilon) |  \bar{c}\gamma^\mu \gamma_5 b |\mathcal{X}^l_{bc}(v',\epsilon') \rangle \nonumber \\
&= & \delta_{il}\sqrt{2M_{cc}M_{bc}}  \Big[  -i  b_0 \epsilon^{\epsilon'\epsilon^* v'\mu} - i b_1 \epsilon^{\epsilon'\epsilon^* v\mu}   \Big]
  \; ,
\end{eqnarray}
where $v^{(\prime)}$, $\epsilon^{(\prime)}$ and $M_{cc(bc)}$ are the 4-velocity, polarization vector and mass of $\mathcal{X}_{cc(bc)}$.
The functions $a_i(q^2)$'s and $b_i(q^2)$'s of the transfer momentum squared $q^2$ are to be determined. 
At the zero recoil (maximal $q^2 = (M_{bc}-M_{cc})^2$) point, they can be obtained by taking the heavy quark limit,
and the results read
\begin{eqnarray}\label{eq:amax}
a_{0,1,2,3}(q^2_{\rm{max}}) = b_{0,1}(q^2_{\rm{max}})  = 1 \; .
\end{eqnarray}
It can be derived in the following way. 
Due to the heavy quark symmetry, the ground state $QQ'$ diquark
is represented by a Lorentz bilinear field (see {\it e.g.} (9) of~\cite{Shi:2020qde})
\begin{eqnarray}
\mathcal{D}_v^{QQ'}(x) = {1+\slashed{v}\over2} [\gamma^\mu A_\mu(x) + i \gamma_5 S(x)]C \; ,
\end{eqnarray}
where $C\equiv i\gamma^0\gamma^2$, the axial-vector field $A_\mu(x)$ annihilates an axial-vector diquark with a polarization vector $\epsilon_\mu$
and the scalar field $S(x)$ annihilates a scalar diquark. All of the color indices are hidden for convenience. 
Its Lorentz transformation property is 
$\mathcal{D}_v(x)\to \mathcal{D}'_{v'}(x') = D(\Lambda)\mathcal{D}_v(\Lambda^{-1}x)D(\Lambda)^{\text{T}}$, 
with $\Lambda$ and $D(\Lambda)$ the Lorentz transformation matrices for Lorentz vectors and Dirac spinors, respectively. 
The conjugate field of $\mathcal{D}_v$ can be introduced as 
$\bar{\mathcal{D}}_v = \gamma^0 \mathcal{D}_v^\dagger \gamma^0$
which transforms as $\bar{\mathcal{D}}_v(x)\to \bar{\mathcal{D}}_{v'}(x') 
= [D(\Lambda)^{-1}]^{\text{T}} \bar{\mathcal{D}}_v(\Lambda^{-1}x)D(\Lambda)^{-1}$. 
Then, we can match the quark transition currents to the corresponding diquark transition currents via
\begin{eqnarray}\label{eq:generalcurrent}
\bar{c}\Gamma b = \text{tr} \left[ L^{\text{T}}\; \bar{\mathcal{D}}^{cQ}_{v'}\; \Gamma \mathcal{D}^{bQ}_v\; \right] \; ,
\end{eqnarray}
which is determined by the heavy quark spin symmetry and Lorentz covariance. The $\Gamma$ matrix 
represents a general $4\times4$ matrix, and only $\gamma^\mu$ and $\gamma^\mu\gamma_5$ are involved in this work. 
The Lorentz bispinor $L$ only depends on $v$ and $v'$. 
The general expression for $L$ with the correct parity and time-reversal properties is
$L = L_0 + L_1 \slashed{v} + L_2 \slashed{v}' + L_3 \slashed{v} \slashed{v}'$, where the coefficients $L_i$ are
functions of $w\equiv v\cdot v'$. The property $\slashed{v}\mathcal{D}_v= \mathcal{D}_v$ together with $v\cdot\epsilon = v'\cdot\epsilon'= 0$
simplifies the current (\ref{eq:generalcurrent}) such that $L^{\text{T}}$ can be replaced by a scalar function $\xi(w) = L_0+L_1+L_2+L_3$. Evaluating the trace 
with $\Gamma= \gamma^\mu,\gamma^\mu\gamma_5$ under (\ref{eq:generalcurrent}) with a proper normalization gives the matrix elements
\begin{eqnarray}
&&{1\over\sqrt{2M_{bc}M_{cc}}}\langle \mathcal{X}_{cc}(v,\epsilon) |  \bar{c}\gamma^\mu b |\mathcal{X}_{bc}(v',\epsilon') \rangle \nonumber \\
&=&  \xi(w) \Big[ - \epsilon^*\cdot\epsilon' v^{\prime\mu} - \epsilon^*\cdot\epsilon' v^\mu
+  \epsilon^*\cdot v' \epsilon^{\prime\mu}  + v\cdot\epsilon' \epsilon^{*\mu} \nonumber \Big] \; ,  \nonumber \\
&&{1\over\sqrt{2M_{bc}M_{cc}}}\langle \mathcal{X}_{cc}(v,\epsilon) |  \bar{c}\gamma^\mu \gamma_5 b |\mathcal{X}_{bc}(v',\epsilon') \rangle \nonumber\\
&=& \xi(w)\Big[  -i \epsilon^{\epsilon'\epsilon^* v'\mu} - i  \epsilon^{\epsilon'\epsilon^* v\mu}   \Big]   \; ,
\end{eqnarray}
where the symmetry factor $\sqrt{2}$ is   due to  the identical $c$ quarks in $\mathcal{X}_{cc}$,
and the masses are introduced  to make $\xi(w)$ dimensionless.
Replacing $\mathcal{X}_{cc}(v,\epsilon) \to \mathcal{X}_{bc}(v',\epsilon')$ and $c\to b$, the above vector-current expression leads to
$\langle \mathcal{X}_{bc}(v',\epsilon') |  \bar{b}\gamma^\mu b |\mathcal{X}_{bc}(v',\epsilon') \rangle /M_{bc}
=  2 \xi(1)\;  v^{\prime\mu}  =  2 v^{\prime\mu}$. 
It determines that $\xi(w) = 1$ at the zero recoil  $w=1$, leading to the final result 
in (\ref{eq:amplitude}) and (\ref{eq:amax}).


In the large-recoil (small-$q^2$) region, the diquark currents are induced by exchanges of hard gluons. 
At the leading order with one hard gluon exchange, a sample Feynman diagram is shown in FIG. \ref{fig:maxrecoil}.  
The hard gluon leads to the large recoil, while the soft gluon exchanges can be absorbed into the initial and final diquark wave functions in the NRQCD framework. 
The NRQCD calculation formulates the diquark currents as nonperturbative matrix elements along with the corresponding Wilson coefficients as
\begin{eqnarray}\label{ffD}
&&a_i[{\chi}_{bc}(v')\to {\chi}_{cc}(v)]  =\sum_{jk} \frac{c^{jk}_i(\mu)}{m_b^{\frac{d_j-4}{2}}m_c^{\frac{d_k-4}{2}}} \nonumber \\
&&\qquad\times\left\langle0\left|{\cal K}'_j(\mu)\right|{\chi}_{bc}(v')\right\rangle \left\langle{\chi}_{cc}(v)\left|{\cal K}_k(\mu)\right|0\right\rangle\;,
\end{eqnarray}
where the ${\cal K}^{(\prime)} (\mu)$ are all possible  independent bilinear combinations of two component operators which
can be power counted by the velocity $v^{(\prime)}$. The $c^{jk}_i(\mu)$ are the short-distance Wilson coefficients which can
be calculated order by order in series of $\alpha_s$. Following an analogous procedure to obtain (4) in \cite{Qiao:2012vt}, we 
calculate the quark-level hard kernel with one hard gluon exchange, convolute it with the diquark nonperturbative 
matrix elements, and obtain the leading-order result
\begin{eqnarray}\label{eq:amin}
a_{2,3}(q^2)&=& {\alpha_s\over 2(1-w)^2\sqrt{w}} {N_c+1\over N_c} {1\over m_c^3}  R_{bc} (0)R_{cc}^*(0)  \; ,\nonumber\\
a_0(q^2) &=& b_0 (q^2) = \bar{\xi}_2 a_{2,3}(q^2) \;, \nonumber \\
a_1(q^2) &=& b_1 (q^2)  = \bar{\xi}_1 a_{2,3}(q^2) \; ,
\end{eqnarray}
where $\bar{\xi}_1\equiv m_b/M_{bc}$, $\bar{\xi}_2\equiv m_c/M_{cc}$, and the number of colors $N_c = 3$.
The diquark wave functions at the origin are defined through the nonperturbative matrix elements
\begin{eqnarray}\label{eq:9}
\varepsilon_{ijk} \frac{\langle 0|\psi_{c,i}^{\text{T}} i\sigma_2{ \vec{\sigma}}\psi_{b,j}|
 {\cal X}_{bc}^k(\vec{\epsilon})\rangle}{\sqrt{2M_{bc}}}  &=& N_c!\; {R_{bc}(0)\over\sqrt{4\pi}} \vec{\epsilon} \; ,\nonumber \\
\varepsilon_{ijk} \frac{\langle {\cal X}_{cc}^k(\vec{\epsilon})|\psi_{c,i}^{\dagger} i{ \vec{\sigma}}\sigma_2\psi^*_{c,j}|
 0\rangle}{\sqrt{4M_{cc}}}  &=& N_c!\; {R_{cc}^* (0)\over \sqrt{4\pi}} \vec{\epsilon}^* \;,
\end{eqnarray}
where $\psi$'s are two-component spinor fields, $\sigma$'s are Pauli matrices and 
$\varepsilon_{ijk}$ is the Levi-Civita symbol in the color space. 
In principle, both the next-to-leading order $\alpha_s$ corrections and the subleading power corrections to the diquark transition amplitudes
can be calculated as the calculation for $B_c\to J/\psi$ in~\cite{Qiao:2012hp,Zhu:2017lqu}. We leave these calculations to future works. 
To obtain the numerical result, it requires the input of the diquark wave functions at the origin $R_{bc,cc}(0)$. They are obtained by solving the nonrelativistic Schr\"odinger equations, with the potential 
$V(r) = -{2\over3}{\alpha_s(\nu_{\text{lat}})\over r} + {c_2 r + c_1\over c_3 r+1} + \sigma r$ \cite{Soto:2020pfa} with 
$\nu_{\text{lat}} = 2.16\ \rm{GeV}\; , \ \sigma = 0.21 \rm{GeV}^2 , 
c_1 = 1.948 \ \rm{GeV}\; ,\ c_2 = 15.782\ \rm{GeV}\; ,\ c_3 = 9.580\ \rm{GeV}$, 
which were fitted from the lattice calculation \cite{Najjar:2009da,Luscher:2002qv}. The quark masses
take values of $m_c = 1.392(11)$ GeV and $m_b = 4.749(18)$ GeV. 
The ground-state solutions to the Schr\"odinger equations give 
${R}_{cc}(0) = (0.66 \pm 0.06) \ \text{GeV}^{3/2}$ and 
${R}_{bc}(0) = (0.87\pm0.09)\ \text{GeV}^{3/2}$. 
The uncertainties were estimated from the differences between the results obtained with the above lattice potentials and 
the Cornell potentials \cite{Bali:2000gf}, though the real uncertainties could be larger.

\begin{figure}\begin{center}
\includegraphics[scale=0.7]{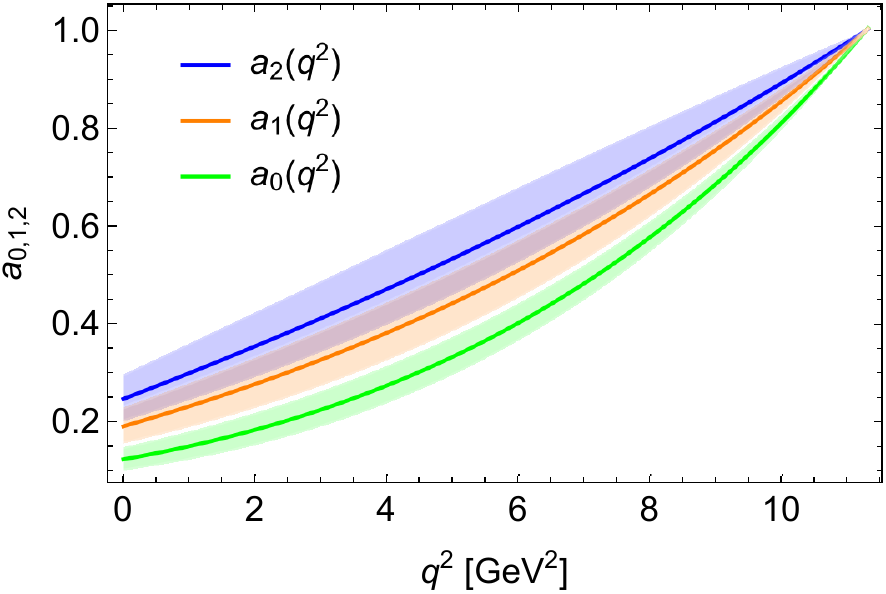}
\caption{The central values (solid curves) and the $\pm1\sigma$ uncertainties (shadow areas) of the numerical results 
for $a_{0,1,2}(q^2)$ appearing in the diquark currents.} \label{fig:ff}
\end{center}\end{figure}

To interpolate the diquark current in the whole range from the above results in the small- and large-recoil regions, 
a simplified $z$-series expansion~\cite{Bourrely:2008za} is adopted with the formulation
\begin{eqnarray}\label{eq:ab}
f(q^2) = {f(0)/ [1-{q^2/ m_{B_c}^2}] }\left[ 1+ b \zeta(q^2) + c\zeta^2(q^2) \right] , 
\end{eqnarray}
for $a_0(=b_0)$, $a_1(=b_1)$ and $a_2(=a_3)$, where 
$\zeta(q^2) = z(q^2) -z(0)$, $z(q^2) = (\sqrt{t_+-q^2} - \sqrt{t_+ - t_0})/( \sqrt{t_+-q^2} + \sqrt{t_+ - t_0})$, 
$t_\pm = (M_{bc} \pm M_{cc})^2$, $t_0 = t_+(1- \sqrt{1-t_-/t_+})$ and 
the free parameters $f(0)$, $b$, and $c$ are to be determined. 
The value of $t_0$ is chosen to minimize $|z|$ to improve the convergence. 
Unlike Ref.~\cite{Bourrely:2008za}, we use $\zeta(q^2)$ instead of $z(q^2)$ 
for the expansion, though these two parameterizations are equivalent. 
The expansion in $\zeta(q^2)$ ensures that $f(0)$ is exactly the value of the form factor at $q^2=0$.
Fitting the points of $q^2=0, 0.1, 0.2, 0.3, 0.4$ GeV$^{2}$ and $q_{\rm max}^2 \approx 11.3$ GeV$^{2}$,
the parameters are extracted as:
$f(0) = 0.247, b=-58.6, c = 238.2~\text{for}\ a_2 \; ;f(0) = 0.124, b=-52.9, c = 1898.2~\text{for}\ a_0 \; ;
f(0)= 0.191, b=-57.0, c = 388.0~\text{for}\ a_1$.
The corresponding results are plotted in FIG.~\ref{fig:ff}, with the uncertainties transferred from 
the diquark wave functions at origin.

Finally, with the numerical results for the diquark currents, the inclusive $\Xi_{bc}\to H_{cc}+X$ decay rate (\ref{eq:lp}) can be calculated.
The leading power free diquark decay rates were calculated by phase space integration of the amplitude 
squares. For example, for the electron channel contribution $\Gamma(\mathcal{X}_{bc}\to \mathcal{X}_{cc}e^-\bar{\nu}_e)$, the amplitude is given by 
the product of ${4G_F/\sqrt{2}} V_{cb}\bar{u}(p_e)\gamma_\mu P_L v(p_\nu)$ and the diquark current
$\langle\mathcal{X}_{cc}|  \bar{c}\gamma^\mu P_L b  |\mathcal{X}_{bc}\rangle$, where $G_F$ is the 
Fermi constant and $V_{qq'}$ is the corresponding Cabbibo-Kobayashi-Maskawa (CKM) matrix element. The calculation is similar for the 
other leptonic channels; for the hadronic channels, the replacement $|V_{cb}|^2 \to |V_{cb}V^*_{UD}|^2 (3 C_1^2 + 2 C_1C_2 + C_2^2)$ 
should be performed at 
the level of amplitude squares with $U=u,c$ and $D=d,s$,
where the Wilson coefficients are defined in Ref.~\cite{Lu:2000em}.
Summing over the contributions from all possible channels with 
$\bar{f}f' =  e\bar\nu, \mu\bar{\nu},\tau\bar{\nu},\bar{u}d,\bar{u}s,\bar{c}d,\bar{c}s$, the numerical result for the inclusive decay rate 
reads 
\begin{eqnarray}
&&\Gamma(\Xi_{bc} \to H_{cc}+X) \nonumber\\
& =& (1.9 \pm0.1\pm0.3 \pm 0.4)\times10^{-13}\ \text{GeV} \;.
\end{eqnarray}
Most numerical inputs have been given previously, except that the Wilson coefficients took values from Ref.~\cite{Lu:2000em} and 
the Fermi constant and the CKM matrix elements took values from Ref.~\cite{ParticleDataGroup:2020ssz}. 
The uncertainties in order are from the quark mass variation, 
the diquark wave functions at the origin, and the scale dependence, respectively. The former two 
were obtained by varying the values of the quark masses and the diquark wave functions at the origin
as listed below (\ref{eq:9}). As for the scale, we chose $\mu =m_b$ 
for the central value calculation, and doubled and halved it for the uncertainty estimation.
In addition, one would expect more uncertainties induced by unknown power corrections. 
The dominant $v^2$ corrections  are expected to potentially modify the result by
$\sim 30\%$~\cite{Zhu:2017lqu}. The decay rate translates to the branching ratios as
\begin{eqnarray}\label{eq:br}
\mathcal{B}(\Xi^{+(0)}_{bc} \to H_{cc}+X) \approx 14\%\; (3\%) \; ,
\end{eqnarray}
where we have taken $\tau(\Xi_{bc}^{+(0)})\approx 508\;(105)$ fs~\cite{Cheng:2019sxr}. 
As analyzed before, the $\Xi^{+,0}_{bc} \to \Xi_{cc}^{++}+X$ branching ratio is approximately 
1/2 of  $\mathcal{B}(\Xi^{+,0}_{bc} \to H_{cc}+X)$.


\textit{Phenomenology.} ---
Based on the $\Xi_{bc} \to \Xi_{cc}^{++}+X$ branching ratio calculated above, as well the information on $\Xi_{bc}$ production
and the $\Xi_{cc}^{++}$ detection efficiency, the number of signal events containing a displaced $\Xi_{cc}^{++}$ can be estimated.

In practice, the inclusive approach to search for $\Xi_{bc}$ depends crucially on the lifetimes of the $\Xi_{bc}$ baryons. 
Only if they fly far enough from the collision vertices before decaying, the displacement of the decay products $\Xi_{cc}^{++}$'s can be clearly distinguished. 
According to the study of $\Xi_{bb} \to B_c^-+X$~\cite{Gershon:2018gda}, with the vertex resolution of the LHCb detector, 
the $\Xi_{bb}$ particles with lifetimes above about 500 fs can lead to displaced $B_c$'s with significantly higher IP values than those of the prompt $B_c$'s. 
In contrast, if their lifetimes are much below 500 fs, the IP values will hardly help separate their decaying $B_c$'s from the prompt ones.
Therefore, we will focus on the $\Xi_{bc}^+$, which is expected to have a sufficiently long lifetime~\cite{Cheng:2019sxr}.

The $\Xi_{bc}$ production cross section at the LHC
has been theoretically evaluated in Refs.~\cite{Zhang:2011hi,Ali:2018xfq}. To reduce 
systematic uncertainties, instead of the direct result for the cross section we adopt the cross section
ratio $\sigma(\Xi_{bc})/\sigma(\Xi_{cc}) \approx 40\%$~\cite{Zhang:2011hi}. 
The signal is determined by a displaced
$\Xi_{cc}^{++}$, so its detection efficiency is expected to be identical to that of a normal $\Xi_{cc}^{++}$,
$\epsilon(\Xi_{cc}^{++})$. With these inputs, the expected signal yield $N_s$ is expressed as
\begin{eqnarray}\label{eq:ns}
N_s &=& N_p(\Xi_{bc}^+) \cdot \mathcal{B}(\Xi_{bc} \to \Xi_{cc}^{++} + X) \cdot \epsilon(\Xi_{cc}^{++}) \nonumber \\
&=& N_d(\Xi_{cc}^{++})\cdot {\sigma(\Xi_{bc})\over \sigma(\Xi_{cc})} \cdot \mathcal{B}(\Xi_{bc}^+ \to \Xi_{cc}^{++} + X) ,
\end{eqnarray}
where $N_{p,d}$ are the number of produced and detected particles. 
Quantitatively, it is expected that LHCb Run 3 will collect approximately $10^4$ $\Xi_{cc}^{++}$'s
through the $\Lambda_c^+K^-\pi^+\pi^+$~\cite{LHCb:2019epo} and $\Xi_c^+\pi^+$~\cite{LHCb:2018pcs} reconstruction. Combining
the inclusive decay branching ratio (\ref{eq:br}) and the $\Xi_{bc}$ production information 
$\sigma(\Xi_{bc})/\sigma(\Xi_{cc})\approx$ 40\%~\cite{Zhang:2011hi}, one finally arrives
at the signal yield at the LHCb Run 3,
$N_s \approx$  300.
In a real measurement, some of these events will get swamped by the background of the primarily produced $\Xi_{cc}^{++}$'s,
if, for example, the $\Xi_{bc}^{+}$'s do not fly far enough before decaying into $\Xi_{cc}^{++}$'s. Although it will lose some efficiencies, 
the $\Xi_{bc}$ discovery will still be hopeful during LHCb Run 3, and will be very promising 
during LHCb Run 4 and at the high-luminosity LHC.

As for the background, a displaced $\Xi_{cc}^{++}$ is also possibly produced from
$B_c^+$ decays, $B_c^+\to \Xi_{cc}^{++} +X$. However, such background is negligible because the 
branching ratio is expected to be tiny due to the phase-space suppression. The dominant quark-level
transition for such decays is $\bar{b}\to c \bar{c} \bar{s}$, so the least massive final state is $\Xi_{cc}^{++} \bar{\Xi}_c^-$, 
with $\sim$0.18 GeV phase space. Considering a similar decay channel $B\to \Lambda_{c} + \bar{\Xi}_c$
with $\sim$0.5 GeV phase space having an $\mathcal{O}(10^{-3})$ branching ratio~\cite{ParticleDataGroup:2020ssz},
the $B_c^+\to \Xi_{cc}^{++} \bar{\Xi}_c^-$ branching ratio is expected to be
even smaller. It also allows decay processes with some other final states
such as $\Xi_{cc} \bar{\Xi}_c\pi$ and $\Xi_{cc} \bar{\Xi}_c^*$,
but all of them are expected to have similar or smaller branching ratios due to even smaller phase spaces compared to $\Xi_{cc}^{++} \bar{\Xi}_c^-$. 
As $B_c$ and $\Xi_{bc}$ have production cross sections of the same order at the LHC~\cite{Ali:2018xfq}, 
the number of the displaced $\Xi_{cc}^{++}$'s produced via $B_c$ decays
is smaller than that of the signal by at least 1 to 2 orders of magnitude. Therefore, this background source can be safely neglected.

With the above analyses/calculations about aspects of the experimental $\Xi_{bc}$ search, we
can conclude that it will be very hopeful to 
discover $\Xi_{bc}$ during LHCb Run 3 via the inclusive approach that we proposed.
The inclusive approach should be more efficient than searches using exclusive decays. 
The exclusive channels induced by $b$ quark decaying typically have 
branching ratios smaller than $\mathcal{B}(\Xi_{bc}\to\Xi_{cc}^{++}+X)$ by more than 1 or 2 orders of magnitude~\cite{Wang:2017mqp}. 
The $bc$ annihilation channels are power suppressed and are thus even rarer. 
The $c$ quark decay channels suffer low reconstruction efficiencies of the $b$-hadrons in their final states~\cite{Han:2021gkl}.

{\it Conclusion.} --- We have proposed that the inclusive $\Xi_{bc}$ decay channel ---or, more explicitly, 
$\Xi_{bc}^+\to\Xi_{cc}^{++}+X$---can be used to search for the $\Xi_{bc}$ baryons, with a very clean and simple signal: a displaced $\Xi_{cc}^{++}$.
By making use of effective theories of QCD, we have calculated its branching ratio at the leading order and found that it is approximately 7\%, 
while the radiative and power corrections are left for future studies. 
Based on the result for the $\Xi_{bc}^+\to\Xi_{cc}^{++}+X$ branching ratio, the $\Xi_{bc}$ production rate, and 
the $\Xi_{cc}^{++}$ detection efficiency extracted from previous experiments, we estimated that LHCb Run 3 can 
accumulate approximately 300 such signal events. The  possible background,
the $B_c^+\to\Xi_{cc}^{++}+X$ decay, has been demonstrated to be negligible. In conclusion, the inclusive
$\Xi_{bc}^+\to\Xi_{cc}^{++}+X$ decay is very likely to serve as the discovery channel for the $\Xi_{bc}$ baryons. 


{\it Acknowledgement.} --- The authors are grateful to Ji-Bo He, Xiang-Peng Wang and Yan-Xi Zhang for useful discussions on the experimental search at the LHC and 
the theoretical framework.
This work is supported by Natural Science Foundation of China under Grants No. 12005068, 11735010, U2032102, 11975112, 12075124 and 12125503, and by the DFG and the NSFC through funds provided to the Sino-German Collaborative Research Center TRR110 “Symmetries and the Emergence of Structure in QCD” (NSFC Grant No. 12070131001, DFG Project-ID 196253076).

\end{document}